\documentclass[twocolumn]{aastex62}

\usepackage{natbib}
\usepackage{amsmath}
\usepackage{mathtools}
\usepackage{graphicx} 
\usepackage{pslatex}
\usepackage{booktabs}
\usepackage{url}

\def \NYVir{NY\ Vir\ }
\def \MJ{M\textsubscript{J}}

\begin{document}
\shorttitle{NY Vir Eclipse timing observations: Two planet solution}
\shortauthors{Song et al.}

\title{AN UPDATED MODEL FOR CIRCUMBINARY PLANETS ORBITING THE sdB BINARY NY VIRGINIS}

\author{Shuo Song}
\affil{University of Iowa \\
Department of Physics and Astronomy, Iowa City, IA 52242, USA} 

\author{Xinyu Mai}
\affil{University of Iowa \\
Department of Physics and Astronomy, Iowa City, IA 52242, USA} 

\author{Robert L. Mutel}
\affil{University of Iowa \\
Department of Physics and Astronomy, Iowa City, IA 52242, USA}

\author{David Pulley}
\affil{British Astronomical Association\\
Burlington House, Piccadilly, London W1J 0DU, UK}

\author{George Faillace}
\affil{British Astronomical Association\\
Burlington House, Piccadilly, London W1J 0DU, UK}

\author{Americo Watkins}
\affil{British Astronomical Association\\
Burlington House, Piccadilly, London W1J 0DU, UK}

\begin{abstract}
We report 18 new primary minima timing observations of the short-period eclipsing binary system NY Virginis. We combined these minima with previously published primary minima to update circumbinary exoplanet models in this system based on O-C timing variations. We performed a non-linear least-squares minimization search using a quadratic ephemeris and either one or two exoplanets.
The only model with an acceptable fit includes a period derivative $\dot{P} = 2.83\times10^{-12}$ and two planets in eccentric orbits $e = 0.15,0.15$  with minimum masses 
2.7 and 5.5 Jovian masses.  Analysis of the orbit stability shows that this solution is stable for at least $10^8$ yr, but a small increase in eccentricity ($e\geq0.20$) for either planet renders the orbits unstable in less than $10^6$ years. A number of model parameters are significantly degenerate, so additional observations are required to determine planetary parameters with high statistical confidence. 
\end{abstract}


\section{Introduction}
  \NYVir (= PG1336-018, GSC 04966-00491; V=13) is an eclipsing binary comprising a hot sub-dwarf B star and a M5 dwarf with an orbital period of 2.42 hours \citep{Kilkenny:1998,Vulkovich:2007}. 
It is a member of the Post-Common Envelope Binary (PCEB) class, in which the primary star, initially in a wide ($\sim$1 AU) binary, has reached the giant branch and the secondary has been engulfed in the giant star's envelope. This results in both energy and angular momentum transfer  onto the outer shell of the giant.  The secondary then spirals inward, so that the binary separation is approximately one solar radius 
\citep{Paczynski:1976,Heber:2009}.  Long-term eclipse  timing observations of these systems are very useful to probe their evolution and dynamics, such as angular momentum transfer and detection of additional components, including exoplanets. This approach for detecting exoplanets has been applied to several short-period eclipsing binaries, including HW~Vir \citep{Lee:2009}, NN~Ser \citep{Beuermann:2010, Marsh:2014}, and NSVS~14256825  \citep{Almeida:2013}.

\cite{Kilkenny:1998} initially fit primary eclipse timing observations of \NYVir with a linear (constant period) ephemeris. However, later observations deviated from this linear prediction significantly. Both \cite{Kilkenny:2011} and \cite{Camurdan:2012} found that the O-C residuals could be fit with a negative quadratic function i.e., a slow decrease in orbital period ($\dot{P}\approx-4\cdot10^{-8}$ day/yr)  which \cite{Camurdan:2012} ascribed to angular momentum loss from the binary system. This loss cannot be caused by  mass transfer between the components, since the eclipsing pair is a detached system and both stars are nearly spherical \citep{Lee:2014}. Rather, the momentum loss is more likely  caused by magnetic braking in the cool secondary \citep{Rappaport:1983}.

As more timing observations became available, the observed O-C timing residuals  began to diverge from a quadratic ephemeris.  \cite{Qian:2012} added a sinusoidal term to the quadratic model, implying the existence of an orbiting circumbinary third body that would cause periodic motion of the binary center of mass with a light travel time modulation of 6.3s. \cite{Lee:2014} observed NY Vir over an additional 39 epochs. They fit two models: A single planet solution with a quadratic term, and a two-planet solution with a constant period. The best fit was a two-planet model, with masses  2.8 and 4.5 Jovian masses and orbital periods  8 and 27 years respectively. 

Unfortunately, these models have recently been invalidated by more recent observations.  \cite{Pulley:2018} reported new eclipse timings for seven sdB binaries, including 15 new times of primary minima for NY Vir. This extended the time baseline by three years to 2017.7. They found that that neither the one-planet model of \cite{Qian:2012} nor the two-planet model of \cite{Lee:2014}  can fit  the full O-C residual time history. They did not attempt a new planetary model fit, arguing instead that more timing observations are required to determine any planetary parameters with high confidence.

Other mechanisms that could account for eclipse timing variations have also been discussed. Both \cite{Qian:2012} and \cite{Lee:2014} considered the possibility that gravitational radiation from the binary may be responsible for the decreasing period, but concluded that the magnitude is two dex lower than the observed effect. Another possibility for the quadratic term is gravitational coupling of the orbit to changes of oblateness of a magnetically active star \citep{Applegate:1992}. \cite{Volschow:2016} considered the Applegate mechanism for a group of sixteen close binaries, including NY Vir. They found that although it could marginally account for observed period variations in several of the eleven listed PCEB binaries, it was insufficient to explain the period variations in the majority of these systems, including NY Vir, where the energy discrepancy was more than a factor of 100 too small.

Finally, not all planetary models that can account for the observed O-C residuals have stable orbits. \citep{Horner:2011} determined that the two-planet solution for the eclipsing polar HU~Aqr proposed by \cite{Qian:2011} is dynamically unstable on a timescale of 5000 yr. Similarly, \citep{Horner:2012} found that the planetary solutions for NN~Ser proposed by 
\cite{Beuermann:2010} are also unstable, although this claim has been disputed \citep{Marsh:2014}.

In this paper, we report new eclipse timing observations of \NYVir that extend the time base to epoch 2018.5. We compare the O-C residuals to several models, including one, two, and multiple planet systems, and evaluate the uniqueness of the derived model parameters. We also present the results of a dynamical model to determine the stability of the derived planetary orbits.

\section{Observations}

We observed \NYVir at primary eclipse on 18 nights between 24 Jan 2018 and 18 April 2018.  The observations were made with the 0.5m telescope at the Iowa Robotic observatory\footnote{\url{http://astro.physics.uiowa.edu/iro}}, located at Winer Observatory\footnote{\url{winer.org}} in southeastern Arizona.  We used a  SBIG 6303e CCD camera (2048x3072 x 12 micron pixels) and a Sloan r$'$ filter.  At each epoch, we took a sequence (typically 20) of 15 sec exposures centered on the predicted time of primary eclipse. 

After the normal CCD image calibration  (dark subtraction, flat fielding), we created  light curves using airmass-corrected differential photometry. 
We found each epoch's time of minimum and associated uncertainty by fitting a Gaussian profile to the primary eclipse light curve using  a nonlinear least-squares (Levenberg-Marquardt ) algorithm.  We then determine the observed mid-eclipse times. Table 1 lists the eclipse times and uncertainties.

\begin{deluxetable}{l}

\tablecaption{NY Vir Primary Mid-Eclipse Times} 
\label{Table_observing}
\tabcolsep=1.8cm
\tablehead
{
\colhead{BJD} }
\startdata
2458142.913026(9)	\\	
2458143.014080(4)  \\
2458143.923178(2)  \\
2458145.943570(3)  	\\
2458146.953672(3) \\
2458147.963801(3)	 \\
2458159.984740(4) 	 \\
2458180.995988(4)   \\
2458190.895673(2)  	\\
2458193.825018(2)   \\
2458201.906339(6)	 \\
2458202.916509(2)  	\\
2458204.936869(4)   \\
2458205.947009(6)   \\
2458206.957149(2)   \\
2458226.756344(1)   \\
2458233.928400(5)   \\ 
2458236.857910(3)    \\
\enddata	
\end{deluxetable}

\section{Additional Data  and Analysis}

We combined the present observations with previously published times of primary minima listed in \cite{Kilkenny:1998},\cite{Kilkenny:2000}, \cite{Kilkenny:2011}, \cite{Camurdan:2012}, \cite{Qian:2012},  \cite{Lee:2014}, \cite{Pulley:2016}, \cite{Pulley:2018} and \cite{Basturk:2018}. In addition, we added several minima found in the AAVSO (aavso.org) and WASP \citep{Butters:2010} public archives.   
We made a few modifications to these data before use:  (a) \cite{Kilkenny:2000} did not publish exact uncertainties of the minima. They reported that the uncertainties are smaller than 0.00005 days. We used 0.00005 days for all their uncertainties.
  (b) The  individual WASP times of minima have large ($\sigma>$10s) uncertainties. These data are concentrated in three seasons, so we took the median O-C value for each season, with an uncertainty given by the standard deviation of all times for that season.
  (c) \cite{Lee:2014}  have several outliers i.e.,  closely-spaced times of minima that differ from by more than three standard deviations from the overall trend. We removed these outliers. 
  
 The aggregate dataset consisted of 104 times of primary minima spanning 23 years, from 1996.4 to 2018.5. 
We calculated a new best-fit linear ephemeris  using all of the available time of minima,
\begin{equation}
JD_{min}(BJD) = 2453174.442647 + 0.1010159677E
\end{equation}
to calculate the O-C residuals and cycle number for all observed primary minima times. The O-C diagram with all of the available data is shown in Figure 1.

It is  clear that the O-C plot is complex, and  excludes both a simple quadratic ephemerides model or strictly periodic variations, as might be expected from e.g., pure apsidal motion or a modulation by a third body in a circular orbit. Hence, we used a family of models that included both a quadratic term and either one or two orbiting bodies in elliptical orbits. The predicted times of minima are 
\begin{equation}
T(E) = T_0 + P_0E + \frac{1}{2}\frac{dP}{dt}E^2 + \tau_1(E) + \tau_2(E)
\end{equation}
where the light travel time delay caused by the reflex motion of the binary by each perturbing body is given by \citep{Irwin:1952, Irwin:1959},
\begin{equation}
\begin{multlined}
\tau_i=  \frac{K_i}{ \sqrt{1-\small[e_icos(\omega_i)\small]^2 } } \cdot \\ 
\bigg[ \frac{1-e_i^2} {1+e_icos(\nu_i)}sin(\nu_i+\omega_i)+e_isin(\omega_i)\bigg]
\end{multlined}
\end{equation}
where the orbit number dependence $E$ is computed by solving Kepler's equation for the eccentric anomaly, followed by the true anomaly $\nu$. 

We used a non-linear least-squares algorithm (Python library LMFIT) to fit the O-C data starting with a simple quadratic model, then adding one or two planets. For the simple quadratic and quadratic plus one planet models, we used a brute-force grid search, followed by a downhill-simplex (Nelder-Mead) chi-square minimization parameter search starting at the brute-force minimum grid point. For the two- and three-planet searches, the number of free parameters was large enough that a brute force  grid  search was computationally prohibitive, so we used a Monte Carlo randomized selection of initial parameter start values and a downhill-simplex minimization search for each set of initial parameters.

\section{Results}
\begin{figure*}[t!]
\includegraphics[scale=0.8]{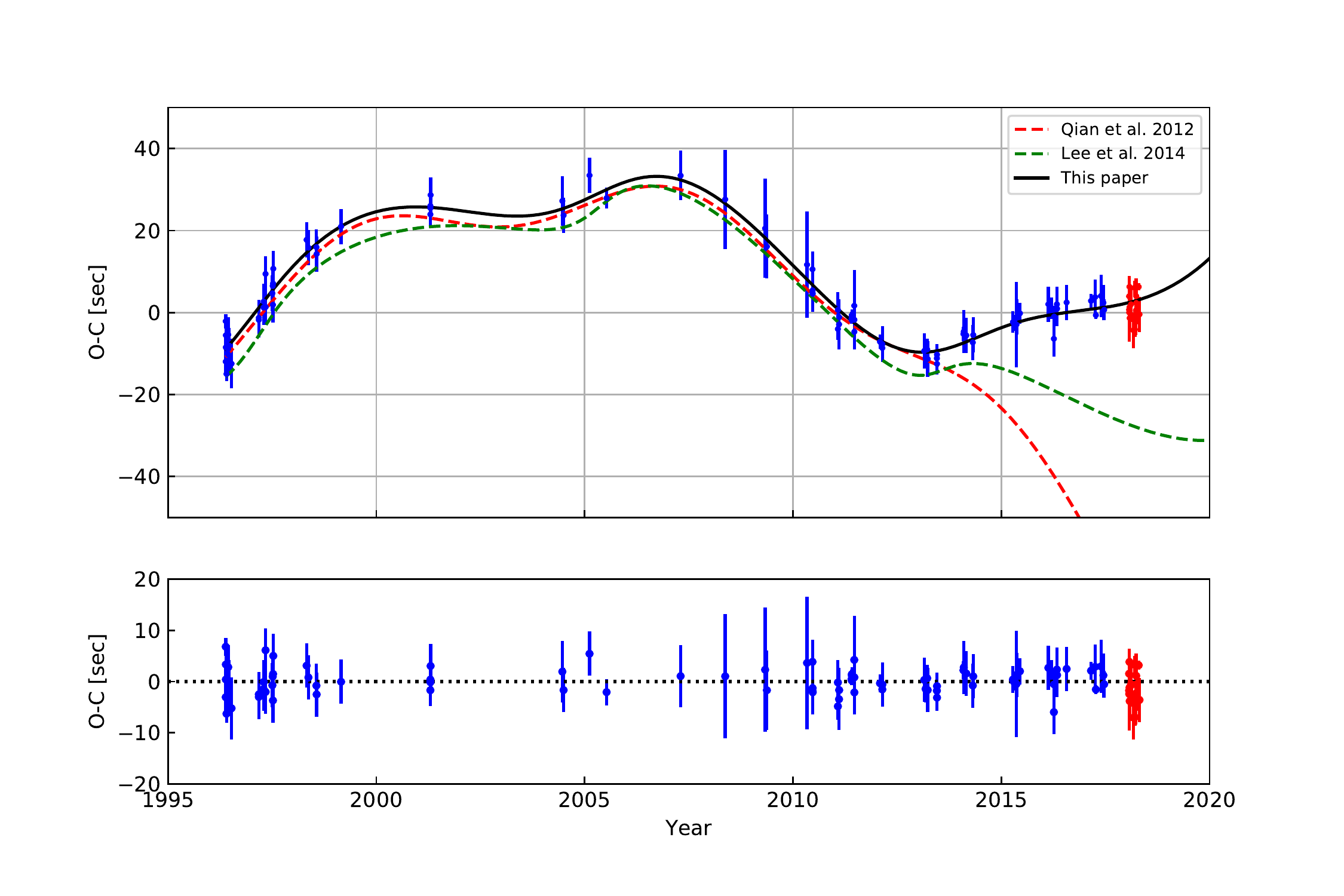}
\caption{[upper plot] NY Virginis O-C times of primary minima computed using the
linear ephemeris given in equation (1). Blue points
are  archival times of minima from the literature (see text), 
while red points are new times of minima from Table 1. Plotted models are: (i) Qian et al. (2012): quadratic ephemeris and one LTT  (red dashed line),
(ii) Lee et al. (2014): linear ephemeris and two LTT's (green dashed line),
(iii) This paper: quadratic ephemeris and two LTT (Table 2, black solid line).
[lower plot] Differences between observed and computed times of primary minima for the  model in this paper (Table 2).
The offsets between earlier data and Qian et al.(2012), Lee et al.(2014) are due to the use of different linear ephemeris.}
\end{figure*}

Figure 1 shows the O-C timing residuals versus epoch computed using the linear ephemeris in equation (1).  Also shown are the quadratic plus one-planet model of \cite{Qian:2012} (red dotted line), the two-planet model of \cite{Lee:2014} (green dotted line), and our best-fit model, a quadratic plus two planet solution (black solid line), summarized in Table 2.  The reduced chi-square of the fit is $\chi^2=1.12$, p =0.17. Both the quadratic only and quadratic plus one planet models, for which a brute-force minimization search was used, resulted in model fits with reduced chi-square greater than 1.9, which can be rejected with high confidence ($p<10^{-6}$). 

Our two-planet solution consists of a quadratic (period derivative) term $dP/dt = (2.83\pm0.28) \cdot 10^{-12}$ s/s and two planets with  minimum masses 2.66$\pm$0.36  \MJ\ and 5.54$\pm$0.28  \MJ,
and orbital periods 8.64$\pm$0.17 yr  and 24.09$\pm$0.65 yr respectively. Note that to convert light travel times to planetary masses, we have assumed a combined stellar binary mass of 0.59 M$_{sun}$ \citep{Vulkovich:2009}.

\begin{deluxetable*}{l r r r}

\tablecaption{Best-fit parameters for the quadratic plus two planet model for NY Vir. 1-$\sigma$ uncertainties are given in parentheses.} 
\label{table_params}
\tabcolsep=0.45cm
\tablehead
{
\colhead{Parameter}   		&\multicolumn{2}{c}{Fitted Values}  & \colhead{Unit} 
}
\startdata
\multicolumn{4}{c}{Inner Binary}														\\
\hline 																		
T$_0$						&\multicolumn{2}{c}{2453174.442647(13)}			&BJD		\\
P$_0$						&\multicolumn{2}{c}{0.1010159677(4)}			&day		\\
dP/dt						&\multicolumn{2}{c}{$2.83(0.25)\cdot10^{-12}$}				&s/s 		\\
\hline
								\multicolumn{4}{c}{Planets}								\\
\hline
						&LTT 1			   &LTT 2					&	units	  \\
P 					        &8.64\ (0.17)	                   &24.09\ (0.65)				 &yr		 \\
$e$						&0.15\ (0.08)	           &0.15\ (0.01)					 &		\\
K				                 &7.6\ (0.7)	           &31.4\ (1.1)             &sec		\\
asin(i)				        &3.55\ (0.01)		   &7.04\ (0.25)				    &AU			\\
T				                &2453472			   &2450031			        &JD			\\
$\omega$			                &348\ (6)			   &320(4)		       &deg			\\
Min. Mass			                &2.66\ (0.26)		    &5.54\ (0.2)		        &\MJ	\\
$\chi_r$					& \multicolumn{2}{c}{1.12 (p = 0.17)} 		    &			\\
\enddata	
\tablenotetext{}{}

\end{deluxetable*}

\section{Discussion}

\subsection{Comparison with previous planetary solutions}
\begin{figure*}[t!]
\centering
\includegraphics[scale=1.2]{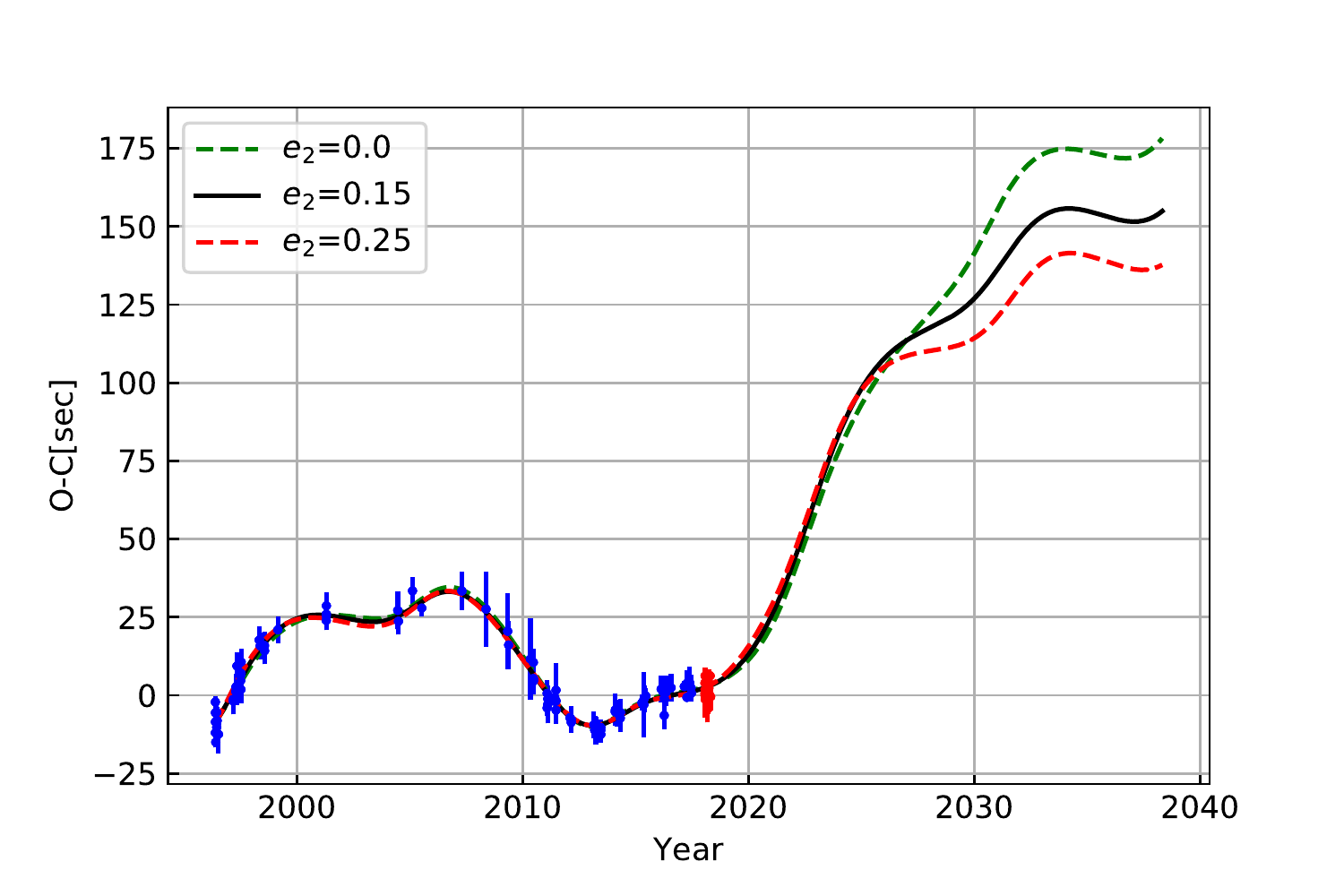}
\caption{\NYVir O-C times of primary minima vs. epoch including model predictions projected to the epoch 2038. All  models comprise a period derivative and two planets.  The parameters have been best-fit to the observed O-C data, but with differing fixed eccentricities for the second planet  (e\textsubscript{2} = 0.0, 0.15, and 0.25 for dashed green, solid  black, and dashed red lines respectively).See text for details.}
\vspace{0.3cm}
\end{figure*}

The planetary model described in this paper most closely resembles the two-planet solution of \cite{Lee:2014}. However, \cite{Lee:2014}'s model had no period derivative term, and does not fit the observed O-C time differences since epoch 2015.0 (cf. Fig.~1).  Likewise, the one-planet plus period-derivative model of \cite{Qian:2012} has a single-planet mass (2.3 \MJ) and period (7.9 yr) similar to our inner planet solution, but has a significantly larger negative period derivative. As with Lee et al. (2014), the \cite{Qian:2012} model does not match recent O-C data. 

\cite{Basturk:2018} recently published a quadratic plus one-planet model that claims to fit all published O-C data through epoch 2017.4 with a remarkably low reduced chi-square value ($\chi^2$= 0.69). However, we could not confirm their results. We were unable to reproduce their published O-C data (Fig. 1) using their linear ephemeris. This is in part because the published linear period (P = 0.101016 $\pm$ 0.000001 day) has insufficient precision: A difference of 0.000001 days over 60,000 periods  changes the O-C plot by 0.06 days, over 100x larger than the O-C departures shown in their Fig. 1. 
We also tried comparing their one-planet model with our O-C data, generated with our linear ephemeris  (Eqn. 1).  This resulted in a very poor fit ($\chi_r^2$ = 3.1, $p \ll 10^{-9}$). We also performed a brute-force chi-squared minimization grid search using their model as a starting point, but were unable to find any one-planet solution with a reduced chi-square less than 1.9 ($p<10^{-6}$). 

Finally, we note that the planetary masses are similar to those derived for \NYVir by \citep{Schleicher:2014} using a second generation common-envelope ejecta model for planet formation (6.2 \MJ\ , 2.5 \MJ\ ), with the larger mass being the outer planet in both cases.

\subsection{How well-determined are the fitted model parameters?}

Although the two-planet model described in this paper fits all available O-C timing observations with an acceptable goodness of fit, it would be naive to accept the fitted model parameters with their associated formal uncertainties as a unique solution, owing to several well-known difficulties associated with non-linear fitting algorithms \citep[e.g.,][]{Transtrum:2010}. These include the often poorly-constrained choice of initial parameter space coordinates,  parameters non-orthogonality (correlation), and overfitting  data by adding additional free parameters (e.g., additional planets) that are unjustified by the dataset. We now discuss each of these briefly in the context of the present model.

As described in section 3, we used a Monte Carlo randomized selection of initial parameter values as a starting point for the simplex chi-squared minimization search in the multi-dimensional parameter space. For each of the 10 independent parameters (4 per planet, plus period and period derivative),  we chose a plausible range, then divided it into 10 uniformly sampled values. This resulted in $10^{10}$ possible starting points, which is clearly computationally impossible to evaluate. Hence we randomly chose a single coordinate as a starting point, then used the Neader-Mead downhill-simplex algorithm to find the nearest local minimum based on reduced chi-square. This procedure was repeated for 50 trials, and the smallest chi-square solution was used as the fitted parameter set.  Although more than 30\% of the trials found the same  minimum, this does not guarantee that the minimum is indeed a global minimum. However, since the reduced chi-square of the chosen fit (1.12, p= 0.17) was already in good agreement, alternate solutions with even lower chi-square values would be only marginally statistically better.

To evaluate possible correlations between parameters, we used the LMFIT Python library implementation of the Nelder-Mead algorithm (method leastsq), which reports the solution cross-correlation matrix. There were about 20 high ($\rho > 0.5$) correlation coefficients, especially between planetary parameters describing the outer planet. For example, the correlation coefficients between the the outer planet eccentricity and 
the period (P\textsubscript{2}), inner planet light-travel time (K\textsubscript{1}), and outer planet argument of periastron ($\omega_2$) all exceeded 0.8. This degeneracy can in principle be broken by additional data, as shown in Fig. 2, where we use four different fixed values of the outer planet's orbital eccentricity, fitting for the other parameters and using the resulting models to predict O-C values 20 years into the future. It is clear that for this example, the degeneracy will not be broken for at least 10 years.

An alternate approach to model fitting, the random-walk Markov Chain Monte Carlo (MCMC) algorithm, is less sensitive to initial parameter guesses, provides a more nuanced description of parameter uncertainty, and has a much higher likelihood of finding a global minimum in chi-square space. MCMC has recently been used to evaluate possible exoplanet parameters derived from light-time variations in several close binary systems e.g.,  NSVS 14256825 \citep{Almeida:2013, Nasiroglu:2017}, NN Ser \citep{Marsh:2014}, and HU Aqr \citep{Gozdziewski:2015}. However, given that the frequentist (least-squares) technique found an acceptable fit at a [possibly local] chi-square minimum and the high degeneracy of the fitted parameters, finding a slightly better fit would not significantly better constrain the exoplanet parameters.

Finally, we consider overfitting, i.e., whether adding additional parameters (e.g., a third planet) is justified. One such test was developed by Schwarz \citep{Schwarz:1978}. In this test, one calculates a dimensionless parameter, the Bayesian information criterion (BIC), for two models with a differing number of parameters. If the difference in BIC is less than 2 \citep{Kass:1995}, the more complex model is not justified. In our case, adding a third planet reduced the BIC from 52.8 (two-planet solution) to 50.4 (three-planet solution), which barely exceeds the unjustified threshold. We conclude that although a three planet model fits the O-C data slightly better,  additional observations will be needed before one could confidently distinguish between these alternatives.

\subsection{Are the planetary orbits stable?}
An obvious question for any multiple planet solution is whether the orbits are dynamically stable over the timescale associated with the system. This in turn depends on whether the planets were  `first generation' i.e., formed coeval with the parent stars or `second generation' i.e., from the ejecta of the common envelope of the close binary \citep{Bear:2014, Hogg:2018}. \cite{Schleicher:2014} considered models for secondary planet formation from disc ejecta for twelve post-common-envelope binaries with evidence of planets from O-C variations, including \NYVir. They found that some planetary systems were more likely first generation, whereas others, including \NYVir, may have formed from common envelope disk material. 
We note that for NY Vir, their two-planet mass estimates 6.2 \MJ\ and 2.4 \MJ\ are in good agreement with our results. 

\cite{Bear:2014} also analyzed the same twelve systems, focusing on angular momentum evolution. They evaluated the ratio of planetary system angular momentum to the  initial binary system angular momentum. For \NYVir, the ratio was 0.05, which is well within the 0.20  limit they adopted for plausible second generation planet formation. If the planets are indeed second-generation, the timescale for planetary orbit stability is likely to be  $t\ll 10^8$ yr, the timescale of post-common envelope evolution \citep[e.g.,][]{Hu:2007}, rather than the age of the constituent stars.

\cite{Lee:2014} analyzed orbit stability of a two-planet solution for \NYVir using a numerical orbit integrator over a timescale of $10^6$ yr. They found that although the system exhibited large-scale instabilities, there were islands of stability in parameter space. In particular, for the outer planet orbit, there was a stable island near semi-major axis $a\sim 7$ AU and eccentricity $e\sim0.15$, which is within one uncertainty of our solution for the outer planet. 

We analyzed orbital stability with Mercury6 program. We used RADAU algorithm with an integration period of 1 day. The results are shown in Figure 3. The solutions given in Table 2 is stable for at least $10^8$ years, which meets the requirement of SG planets. We did more analysis with different sets of parameters. Since the mass and semi-major axis of the planets of different solutions of NY Vir are similar, the eccentricities of the two planets determine that stability of the system. More specifically, when eccentricities of both planets reach 0.2, or when eccentricity of one of the planets reaches 0.3, this system becomes unstable. 

\begin{figure*}[t!]
\includegraphics[scale=0.8]{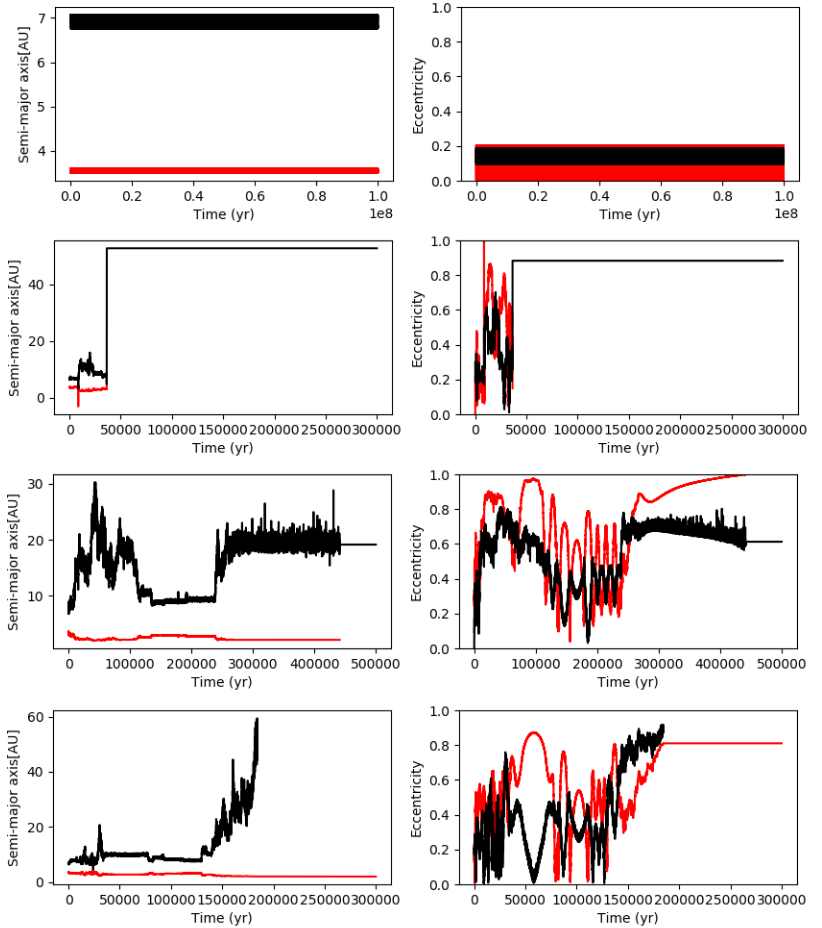}
\caption{Stability analysis using the Mercury6 program with the  RADAU algorithm. Red lines represent the inner planet, and the black lines represent the outer planet. Figures on the left show how the semi-major axis of each planet varies with time. Figures on the right show how the eccentricity of each planet varies with time. From top to bottom are: (i) Solutions of this paper ($e_1$=0.15, $e_2$=0.15), (ii) Solution with $e_1$=0.3, $e_2$=0.0, (iii) Solution with $e_1$=0.3, $e_2$=0.0, (iv) Solution with $e_1$=0.2, $e_2$=0.2.}
\end{figure*}

\section{Summary}
We report new eclipse timing observations of the  short period sdB binary \NYVir. The new times of primary minimum do not agree with predictions from previously published one and two planet models calculated from O-C deviations from a linear ephemeris. We found a new stable two-planet plus a period derivative solution that is in agreement with all historical O-C timing data. However, we find that most model parameters (planetary masses, orbital parameters, and period derivative) are highly degenerate, with large cross-correlations between parameters.  The degeneracy can be broken by additional timing data, but the required timescale for breaking the degeneracies is at least a decade. 

\section{Acknowledgments}
This research made use of the LMFIT non-linear least-square minimization Python library, and Astropy\footnote{http://www.astropy.org} a community-developed core Python package for Astronomy. We also made used of the Mercury orbit stability analysis program \cite{Chambers:1999}.The Iowa Robotic Observatory is funded by the College of Liberal Arts at the University of Iowa. The project was also funded in part by NSF award 1517412. 

\bibliography{NY_Vir_3-14}

\begin{thebibliography}{36}
\expandafter\ifx\csname natexlab\endcsname\relax\def\natexlab#1{#1}\fi

\bibitem[{Almeida {et~al.}(2013)Almeida, Jablonski, \&
  Rodrigues}]{Almeida:2013}
Almeida, L., Jablonski, F., \& Rodrigues, C.~V. 2013, The Astrophysical
  Journal, 766

\bibitem[{Applegate(1992)}]{Applegate:1992}
Applegate, J.~H. 1992, \apj, 385, 621

\bibitem[{Ba{\c s}t{\"u}rk \& Esmer(2018)}]{Basturk:2018}
Ba{\c s}t{\"u}rk, {\"O}., \& Esmer, E.~M. 2018, Open Astronomy, 27, 14

\bibitem[{Bear \& Soker(2014)}]{Bear:2014}
Bear, E., \& Soker, N. 2014, Monthly Notices of the Royal Astronomical Society,
  444, 1698

\bibitem[{Beuermann {et~al.}(2010)Beuermann, Hessman, Dreizler, Marsh, Parsons,
  Winget, Miller, Schreiber, Kley, Dhillon, Littlefair, Copperwheat, \&
  Hermes}]{Beuermann:2010}
Beuermann, K., Hessman, F.~V., Dreizler, S., Marsh, T.~R., Parsons, S.~G.,
  Winget, D.~E., Miller, G.~F., Schreiber, M.~R., Kley, W., Dhillon, V.~S.,
  Littlefair, S.~P., Copperwheat, C.~M., \& Hermes, J.~J. 2010, \aap, 521, L60

\bibitem[{Butters {et~al.}(2010)Butters, West, Anderson, Collier~Cameron,
  Clarkson, Enoch, Haswell, Hellier, Horne, Joshi, Kane, Lister, Maxted,
  Parley, Pollacco, Smalley, Street, Todd, Wheatley, \& Wilson}]{Butters:2010}
Butters, O.~W., West, R.~G., Anderson, D.~R., Collier~Cameron, A., Clarkson,
  W.~I., Enoch, B., Haswell, C.~A., Hellier, C., Horne, K., Joshi, Y., Kane,
  S.~R., Lister, T.~A., Maxted, P.~F.~L., Parley, N., Pollacco, D., Smalley,
  B., Street, R.~A., Todd, I., Wheatley, P.~J., \& Wilson, D.~M. 2010, \aap,
  520, L10

\bibitem[{Caceres {et~al.}(2014)Caceres, Copperwheat, Breedt, Bours, Schreiber,
  Parsons, Littlefair, Dhillon, \& Marsh}]{Marsh:2014}
Caceres, C., Copperwheat, C.~M., Breedt, E., Bours, M. C.~P., Schreiber, M.~R.,
  Parsons, S.~G., Littlefair, S.~P., Dhillon, V.~S., \& Marsh, T.~R. 2014,
  Monthly Notices of the Royal Astronomical Society, 437, 475

\bibitem[{{\c C}amurdan {et~al.}(2012){\c C}amurdan, Zengin~{\c C}amurdan, \&
  {\.I}bano{\v g}lu}]{Camurdan:2012}
{\c C}amurdan, C.~M., Zengin~{\c C}amurdan, D., \& {\.I}bano{\v g}lu, C. 2012,
  \na, 17, 325

\bibitem[{Chambers(1999)}]{Chambers:1999}
Chambers, J.~E. 1999, \mnras, 304, 793

\bibitem[{Go{\'z}dziewski {et~al.}(2015)Go{\'z}dziewski, S{\l}owikowska,
  Dimitrov, Krzeszowski, {\.Z}ejmo, Kanbach, Burwitz, Rau, Irawati, Richichi,
  Gawro{\'n}ski, Nowak, Nasiroglu, \& Kubicki}]{Gozdziewski:2015}
Go{\'z}dziewski, K., S{\l}owikowska, A., Dimitrov, D., Krzeszowski, K.,
  {\.Z}ejmo, M., Kanbach, G., Burwitz, V., Rau, A., Irawati, P., Richichi, A.,
  Gawro{\'n}ski, M., Nowak, G., Nasiroglu, I., \& Kubicki, D. 2015, \mnras,
  448, 1118

\bibitem[{Heber(2009)}]{Heber:2009}
Heber, U. 2009, Annual Review of Astronomy and Astrophysics, 47, 211

\bibitem[{Hu {et~al.}(2007)Hu, Nelemans, \O{}stensen, Aerts, Vuckovi\'{}c, \&
  Groot}]{Hu:2007}
Hu, H., Nelemans, G., \O{}stensen, R., Aerts, C., Vuckovi\'{}c, M., \& Groot,
  P.~J. 2007, A\&A, 473, 569

\bibitem[{Irwin(1952)}]{Irwin:1952}
Irwin, J.~B. 1952, \apj, 116, 218

\bibitem[{Irwin(1959)}]{Irwin:1959}
---. 1959, \aj, 64, 149

\bibitem[{Kass \& Raftery(1995)}]{Kass:1995}
Kass, R.~E., \& Raftery, A.~E. 1995, Journal of the American Statistical
  Association, 90, 773

\bibitem[{Kilkenny(2011)}]{Kilkenny:2011}
Kilkenny, D. 2011, Monthly Notices of the Royal Astronomical Society, 412, 487

\bibitem[{Kilkenny {et~al.}(2000)Kilkenny, Keuris, Marang, Roberts, van Wyk, \&
  Ogloza}]{Kilkenny:2000}
Kilkenny, D., Keuris, S., Marang, F., Roberts, G., van Wyk, F., \& Ogloza, W.
  2000, The Observatory, 120, 48

\bibitem[{Kilkenny {et~al.}(1998)Kilkenny, Koen, O'Donoghue, van Wyk, \&
  Lynas-Gray}]{Kilkenny:1998}
Kilkenny, D., Koen, C., O'Donoghue, D., van Wyk, F., \& Lynas-Gray, A.~E. 1998,
  Monthly Notices of the Royal Astronomical Society, 296, 329

\bibitem[{Lee {et~al.}(2009)Lee, Kim, Kim, Koch, Lee, Kim, \& Park}]{Lee:2009}
Lee, J.~W., Kim, S.-L., Kim, C.-H., Koch, R.~H., Lee, C.-U., Kim, H.-I., \&
  Park, J.-H. 2009, The Astronomical Journal, 137, 3181

\bibitem[{Lee {et~al.}(2014)Lee, Youn, \& Han}]{Lee:2014}
Lee, Jae Wooand~Hinse, T.~C., Youn, J.-H., \& Han, W. 2014, Monthly Notices of
  the Royal Astronomical Society, 445, 2331

\bibitem[{Nasiroglu {et~al.}(2017)Nasiroglu, Go{\'z}dziewski, S{\l}owikowska,
  Krzeszowski, {\.Z}ejmo, Zola, Er, Og{\l}oza, Dr{\'o}{\.z}d{\.z},
  Koziel-Wierzbowska, Debski, \& Karaman}]{Nasiroglu:2017}
Nasiroglu, I., Go{\'z}dziewski, K., S{\l}owikowska, A., Krzeszowski, K.,
  {\.Z}ejmo, M., Zola, S., Er, H., Og{\l}oza, W., Dr{\'o}{\.z}d{\.z}, M.,
  Koziel-Wierzbowska, D., Debski, B., \& Karaman, N. 2017, \aj, 153, 137

\bibitem[{Nixon {et~al.}(2018)Nixon, Wynn, \& Hogg}]{Hogg:2018}
Nixon, C., Wynn, G.~A., \& Hogg, M.~A. 2018, Monthly Notices of the Royal
  Astronomical Society, 479, 4486

\bibitem[{Paczynski(1976)}]{Paczynski:1976}
Paczynski, B. 1976, Symposium - International Astronomical Union, 73, 75

\bibitem[{Pulley {et~al.}(2016)Pulley, Faillace, Smith, \&
  Watkins}]{Pulley:2016}
Pulley, D., Faillace, G., Smith, D., \& Watkins, A. 2016, Journal of the
  British Astronomical Association, 126, 249

\bibitem[{Pulley {et~al.}(2018)Pulley, Faillace, Smith, Watkins, \& von
  Harrach}]{Pulley:2018}
Pulley, D., Faillace, G., Smith, D., Watkins, A., \& von Harrach, S. 2018,
  A\&A, 611, A48

\bibitem[{Qian {et~al.}(2012)Qian, Zhu, Dai, Fern{\'{a}}ndez-Laj{\'{u}}s,
  Xiang, \& He}]{Qian:2012}
Qian, S.-B., Zhu, L.-Y., Dai, Z.-B., Fern{\'{a}}ndez-Laj{\'{u}}s, E., Xiang,
  F.-Y., \& He, J.-J. 2012, The Astrophysical Journal, 745, L23

\bibitem[{Rappaport {et~al.}(1983)Rappaport, Verbunt, \& Joss}]{Rappaport:1983}
Rappaport, S., Verbunt, F., \& Joss, P.~C. 1983, \apj, 275, 713

\bibitem[{Schleicher \& Dreizler(2014)}]{Schleicher:2014}
Schleicher, D. R.~G., \& Dreizler, S. 2014, A\&A, 563, A61

\bibitem[{Schwarz(1978)}]{Schwarz:1978}
Schwarz, G. 1978, Ann. Statist., 6, 461

\bibitem[{Tinney {et~al.}(2012)Tinney, Wittenmyer, Horner, \&
  Hinse}]{Horner:2012}
Tinney, C.~G., Wittenmyer, R.~A., Horner, J., \& Hinse, T.~C. 2012, Monthly
  Notices of the Royal Astronomical Society, 425, 749

\bibitem[{Tinney {et~al.}(2011)Tinney, Wittenmyer, Horner, \&
  Marshall}]{Horner:2011}
Tinney, C.~G., Wittenmyer, R.~A., Horner, J., \& Marshall, J.~P. 2011, Monthly
  Notices of the Royal Astronomical Society: Letters, 416, L11

\bibitem[{Transtrum {et~al.}(2010)Transtrum, Machta, \&
  Sethna}]{Transtrum:2010}
Transtrum, M.~K., Machta, B.~B., \& Sethna, J.~P. 2010, Physical Review
  Letters, 104, 060201

\bibitem[{V\"olschow {et~al.}(2016)V\"olschow, Schleicher, Perdelwitz, \&
  Banerjee}]{Volschow:2016}
V\"olschow, M., Schleicher, D. R.~G., Perdelwitz, V., \& Banerjee, R. 2016,
  A\&A, 587, A34

\bibitem[{Vuckovi\'c {et~al.}(2007)Vuckovi\'c, Aerts, \O{}stensen, Nelemans,
  Hu, Jeffery, Dhillon, \& Marsh}]{Vulkovich:2007}
Vuckovi\'c, M., Aerts, C., \O{}stensen, R., Nelemans, G., Hu, H., Jeffery,
  C.~S., Dhillon, V.~S., \& Marsh, T.~R. 2007, A\&A, 471, 605

\bibitem[{Vuckovi\'c {et~al.}(2009)Vuckovi\'c, \O{}stensen, Aerts, Telting,
  Heber, \& Oreiro}]{Vulkovich:2009}
Vuckovi\'c, M., \O{}stensen, R.~H., Aerts, C., Telting, J.~H., Heber, U., \&
  Oreiro, R. 2009, A\&A, 505, 239

\bibitem[{Zhao {et~al.}(2011)Zhao, He, Liao, Dai, Zhang, Li, Liu, Li, Zhu, \&
  Qian}]{Qian:2011}
Zhao, E.-G., He, J.-J., Liao, W.-P., Dai, Z.-B., Zhang, J., Li, K., Liu, L.,
  Li, L.-J., Zhu, L.-Y., \& Qian, S.-B. 2011, Monthly Notices of the Royal
  Astronomical Society: Letters, 414, L16

\end{thebibliography}

\end{document}